# A new method to probe conducting filaments in $MoS_2$-based memristors

Pierre Trousset[1], Lucie Le Van-Jodin[1], Bruno Reig[1], Clotilde Ligaud[1], Thomas Jalabert[1], Hanako Okuno[2], Le Van-Hoan[1], Paul Brunet[1], Stéphane Cadot[1] and Matthieu Jamet[3]

[1] University Grenoble Alpes, CEA, Leti, F-38000 Grenoble, France

[2] University Grenoble Alpes, CEA, IRIG, MEM, F38000 Grenoble, France

[3] University Grenoble Alpes, CEA, CNRS, IRIG-Spintec, F-38000 Grenoble, France

Two-dimensional (2D) transition metal dichalcogenides (TMDs), such as molybdenum disulfide ($MoS_2$), are emerging as promising materials for next-generation electronic devices. They have proved to be serious candidates for integration with memristors in non-volatile memory and radio frequency (RF) applications. However, the physical mechanisms behind their resistive switching, particularly the formation and resorption of conducting filaments, remain unclear. In this study, we present a novel mechanical exfoliation technique that selectively removes the top metallic electrode from $MoS_2$-based memristors by exploiting the weak van der Waals interaction between $MoS_2$ and the top electrode. This method enables direct and multi-scale characterization of the $MoS_2$ surface in different states (initial, ON and OFF) using Kelvin Probe Force Microscopy (KPFM) and Raman spectroscopy mapping. To complete this study, cross-sectional Transmission Electron Microscopy (TEM) was also performed in different conductive states. Our results reveal that the conducting filament is formed by metallic atom migration from the top electrode into the $MoS_2$ layer. Additionally, we demonstrate that the choice of metallic electrodes (gold vs. nickel) significantly impacts the switching behavior due to differences in adsorption and diffusion energies. This work not only clarifies the filament formation mechanism and introduces a reproducible approach for in-operando characterization but also represents a real progress in the understanding and optimization of 2D material-based memristors.

## Introduction

Today, great research efforts are dedicated to the development of new materials to achieve even smaller and ultralow power consumption electronic components. In this context, 2D semiconducting materials are among the most promising candidates to replace silicon in several applications by keeping exceptional electronic properties down to the monolayer form (<1 nm). Among 2D materials, the Transition Metal Dichalcogenides (TMDs) family is of particular interest and extensively studied [1]. Their general formula is $MX_2$ where M is a transition metal (M= Mo, W, Pt …) and X a chalcogen (X=Se, S, Te…). The first electronic devices incorporating 2D materials were made of micrometric flakes transferred by mechanical exfoliation from the bulk material with scotch tape [2]. More recently, wafer-scale growth methods have been developed using atomic layer deposition (ALD), metalorganic chemical vapor deposition (MOCVD) and molecular beam epitaxy (MBE) [3], [4], [5], [6], facilitating their integration in microelectronic devices.

They are many applications for TMDs. They have already been integrated successfully in flexible devices [7], transistors [8], [9], [10], [11], photodetectors[12], [13], [14], sensors[15], [16], RF switches devices [17], [18], [19] , multi-level memory [20], [21], [22] and non-volatile memory devices [23], [24], [25].

In particular, recent research highlights the potential of 2D materials in RF switches [19] and non-volatile memories [17]. In these devices, memristors were designed with a $MoS_2$ film sandwiched between two metallic electrodes. Promising performances were reached with an endurance of 150 cycles [26] or a cut-off frequency of 70 THz [17]. The physical mechanism responsible for the memristive effect in this type of devices was studied, either by simulations [27], [28], [29], [30], or experimentally [31], [32]. The formation of a conductive pathway appears to be accepted by most of the studies. However, the nature of the filament and the precise mechanisms involved for its formation have not been fully elucidated yet and several hypotheses have been proposed. Tang et. al [31] highlighted the role of $MoS_2$ grain boundaries and suggested the migration of sulfur vacancies into the grain boundaries to be the switching mechanism in $MoS_2$ memristors. Another hypothesis is the migration of metallic ions from the electrode into sulfur vacancies [27], [28], [33]. However, most of these works were performed without direct observation of the conducting path or by isolating one specific microscopic mechanism, preventing another possibility. Only few studies showed the filament, most of the time by transmission electronic microscopy (TEM) cross view [32]. They show a metallic filament through the 2D materials. Interesting results have also been obtained by cycling devices with an AFM tip [33].

In this work, we fabricate and characterize $MoS_2$ memristors in order to gain better insight into the switching mechanism by both electrical measurements and the direct observation of the conducting path inside the device. For this purpose, using a CMOS compatible process on large area, we obtain many devices and then we can statistically compare the performances of memristors. By changing the nature of the metallic electrodes, it allowed us to study the role of the electrodes in the filament formation. The key progress relies on the development of a new protocol taking advantage of the weak van der Waals interaction between $MoS_2$ and the metallic electrode, which allows direct observation of the $MoS_2$/metal interface after fabrication and preparation in ON and OFF states. By combining Kelvin probe force microscopy (KPFM), Raman spectroscopy and cross section TEM, we eventually image the conducting channel and attribute it to a full migration of metallic atoms into $MoS_2$. In this work, we thus present an original method opening a new route to visualize and fully characterize from the micrometer to the atomic scale individual conducting paths in functional 2D material based memristors.

## Device fabrication

In order to study the impact of the nature of metallic electrodes on the memristive effect, different devices with different electrodes are fabricated, all with a tri-layer of $MoS_2$ grown by ALD [3] with a grain size of approximately 10 nm (see Fig. 1(a)).

Fig. 1(b) shows the stacking of a $MoS_2$ memristor (Fig. 1(c)) made of $MoS_2$ sandwiched between a bottom and a top electrode. The whole process is described in Ref. [24]. Bottom electrodes made of nickel or platinum, are fabricated on 200 mm wafer (Fig. 1(d)). They are planarized to enable a great conformity and a good electrical contact with $MoS_2$. Then, gold or nickel is directly evaporated onto the $MoS_2$ to ensure a good interface between the top electrode and $MoS_2$ and not to damage the $MoS_2$ film during transfer. Then, polymethyl methacrylate (PMMA) is spin coated onto the $MoS_2$/metal film and the entire stack is transferred onto a chip containing the planarized bottom electrodes using a wet transfer method [34]. After the transfer step, $MoS_2$ is patterned using deep UV lithography. 100 nm of gold are deposited by evaporation and patterned by a lift-off process to design the device and complete the top electrode. Fig. 1(e) shows a TEM cross section view of a device after the whole fabrication process completed.

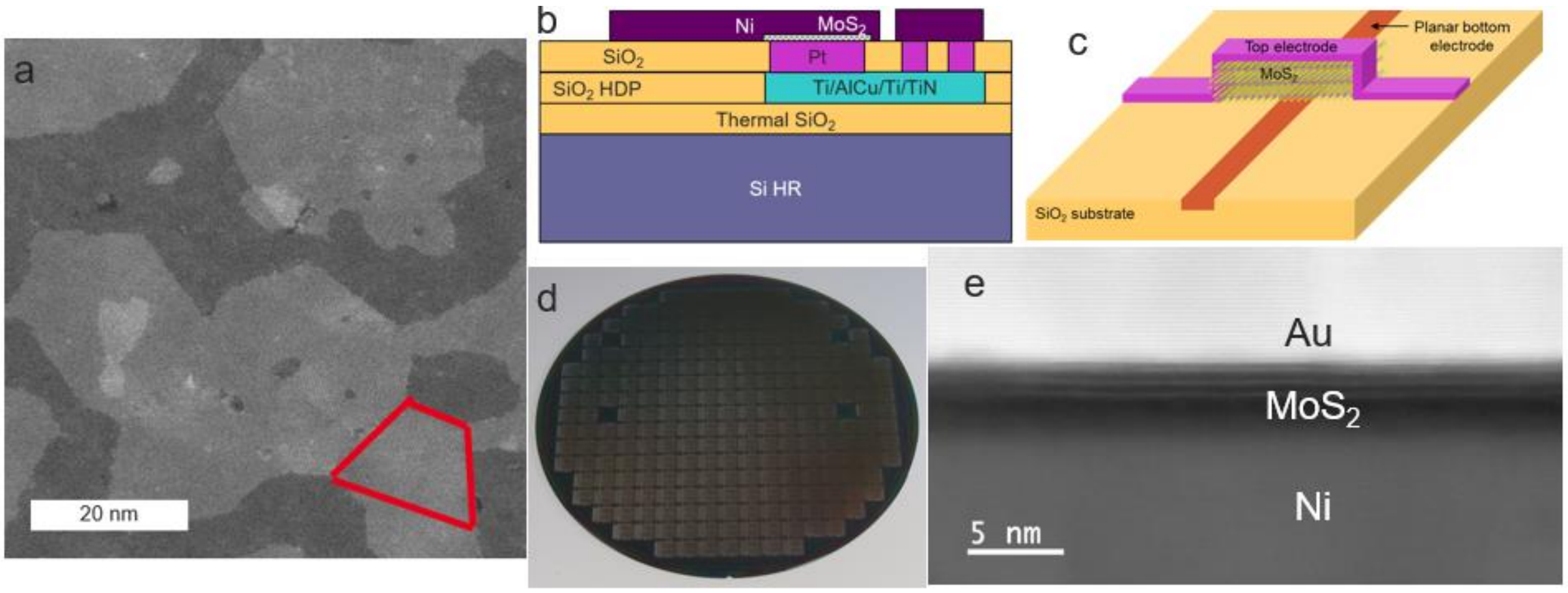

*Figure 1: (a) TEM view of the tri-layer of $MoS_2$ grown by ALD. The area delimited in red corresponds to an individual $MoS_2$ grain. (b) Cross section schematics of a $MoS_2$ memristor with a platinum bottom electrode and a nickel top electrode. (c) View in perspective of a $MoS_2$ memristor. (d) Optical image of a 200 mm wafer with patterned bottom electrodes. (e) Cross section TEM view of a $MoS_2$ memristor with a nickel bottom electrode and a gold top electrode.*

## Electrical measurements

We first study the memristor cycling endurance following the method of Ref. [24]. For this purpose, the top electrode is grounded and the positive or negative voltage is applied to the bottom electrode. A positive voltage is first applied to achieve the forming operation corresponding to the transition from the initial insulating state to the first conducting state. Then, a negative voltage is applied to switch from the conducting state or ON state to the OFF state. This operation is called the reset operation. A positive voltage is then applied to switch from the OFF to the ON state, this operation is called the set operation. We consider a full cycle completed when a reset operation and a set operation are successfully achieved, regardless the forming operation. A compliance current of 0.01 A is set for the forming and set operations to limit the current in the device and facilitate the reset operation. There is no compliance current for the reset operation. At the end of each set or reset operation, a low voltage of 0.1 V is applied to measure the current in ON and OFF states. The key parameters to study are: the cycling endurance that corresponds to the number of cycles a device can achieve; the switching voltage that corresponds to the voltage when the conducting state changes in both set and reset operations; the $I_{ON}/I_{OFF}$ ratio that corresponds to the conductivity contrast between ON and OFF states.

To study the impact of the nature of the electrode, we cycled devices with two different stacks, Ni/$MoS_2$/Au and Pt/$MoS_2$/Ni. Fig. 2(a) and 2(b) (resp. 2(c) and 2(d)) show the cycles and the ON and OFF states for each cycle of a Ni/$MoS_2$/Au (resp. Pt/$MoS_2$/Ni)) device.

Regarding the cycle-to-cycle variability, both stacks exhibit the same behavior. The OFF state and switching voltage for the set operation show a significant variability depending on the cycle. The ON state and switching voltage of the reset operation are quite stable. Moreover, the $I_{ON}/I_{OFF}$ ratio is around 100 for both stacks. Nevertheless, some differences can be observed. Fig. 4(e) compares the set and reset switching voltages which are lower for Ni/$MoS_2$/Au devices than for Pt/$MoS_2$/Ni devices. The mean set voltage value for Ni/$MoS_2$/Au stack is 0.6 V while it is 0.79 V for Pt/$MoS_2$/Ni. In the same way, the reset value is -0.4 V lower for Ni/$MoS_2$/Au and -0.59 V for Pt/$MoS_2$/Ni. About 50 Ni/$MoS_2$/Au devices could be measured and 33 % of them completed at least one cycle. The highest cycling endurance recorded is 24 cycles. Among the 20 Pt/$MoS_2$/Ni devices measured, one did not switch and 3 of them show more than 100 cycles. Considering all the devices, the Pt/$MoS_2$/Ni device endurance is 10 times higher than that of Ni/$MoS_2$/Au devices. All these observations are explained in the following using a multiscale analysis of individual conducting paths made possible by the mechanical exfoliation of the top metallic electrode.

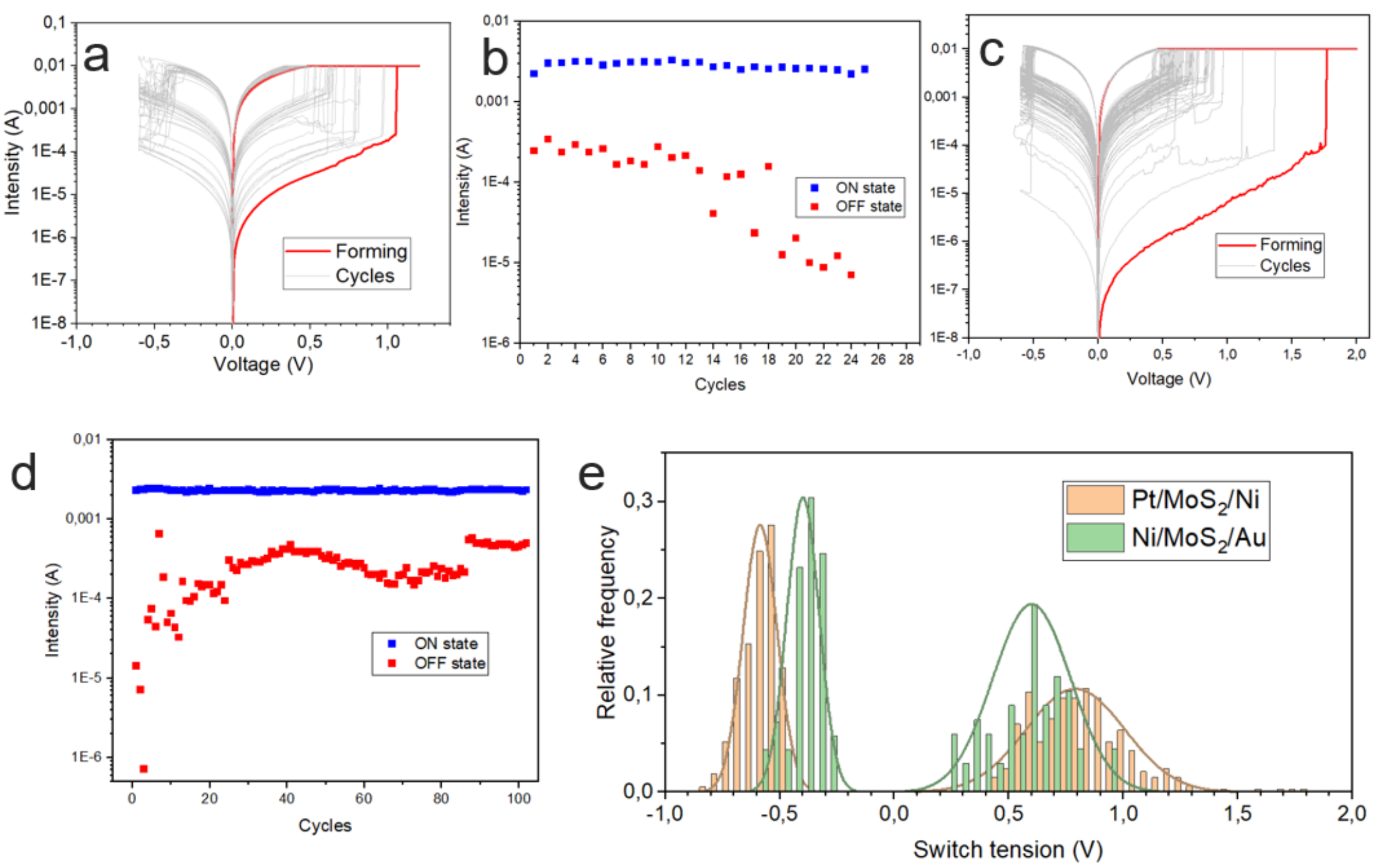

*Figure 2: (a) Ni/$MoS_2$/Au device cycling curves (b) Current intensity in the ON and OFF states of the same Ni/$MoS_2$/Au device as a function of the number of cycles (c) Typical Pt/$MoS_2$/Ni device cycling curves (d) Current intensity in the ON and OFF states of the same Pt/$MoS_2$/Ni device as a function of the number of cycles.(e) Switching voltage distribution for Pt/$MoS_2$/Ni (orange) and Ni/$MoS_2$/Au (green) devices curve for the initial state of devices.*

## Top electrode exfoliation

In order to understand the switching mechanism in memristor devices, it is essential to observe the $MoS_2$ layer in different states: initial state, ON state and OFF state. However, due to the top metallic electrode, $MoS_2$ is not accessible with surface characterization tools. To overcome this obstacle, we develop a mechanical exfoliation process of the top electrode. The objective is to take advantage of the weak van der Waals interaction between $MoS_2$ and the top electrode to exfoliate only the top metallic electrode and leave the $MoS_2$ on the sample.

For this purpose, a 200 nm-thick nickel film is deposited onto the devices by electron beam evaporation. This layer has an intrinsic compressive stress and a good adhesion to the gold top electrode layer. Then, a piece of adhesive tape is placed on the nickel film and peeled off. Due to the stress in the nickel layer, a spalling effect occurs and the top layer splits at the weakest interface [35]. In that case, the weakest interface is the one between the gold electrode and $MoS_2$ corresponding to a van der Waals interface. The top electrode is removed, revealing the $MoS_2$ surface, now accessible to direct characterization. Figure 3 (a) and (b) respectively show an illustration of the device before and after the exfoliation in cross section. The top electrode is removed while the $MoS_2$ remains in the device. Figure 3 (c) and (d) show respectively the optical microscopy image in top view of a device before and after top electrode exfoliation. We can observe that only the gold film above $MoS_2$ is peeled off, giving a direct access to the $MoS_2$ using electro-chemical characterization tools. The $MoS_2$ layer is not visible optically but Raman spectroscopy confirms its presence. The small grey point at the center corresponds to the bottom electrode.

This exfoliation method works only with Ni/$MoS_2$/Au devices and all the following characterizations are made on this stack.

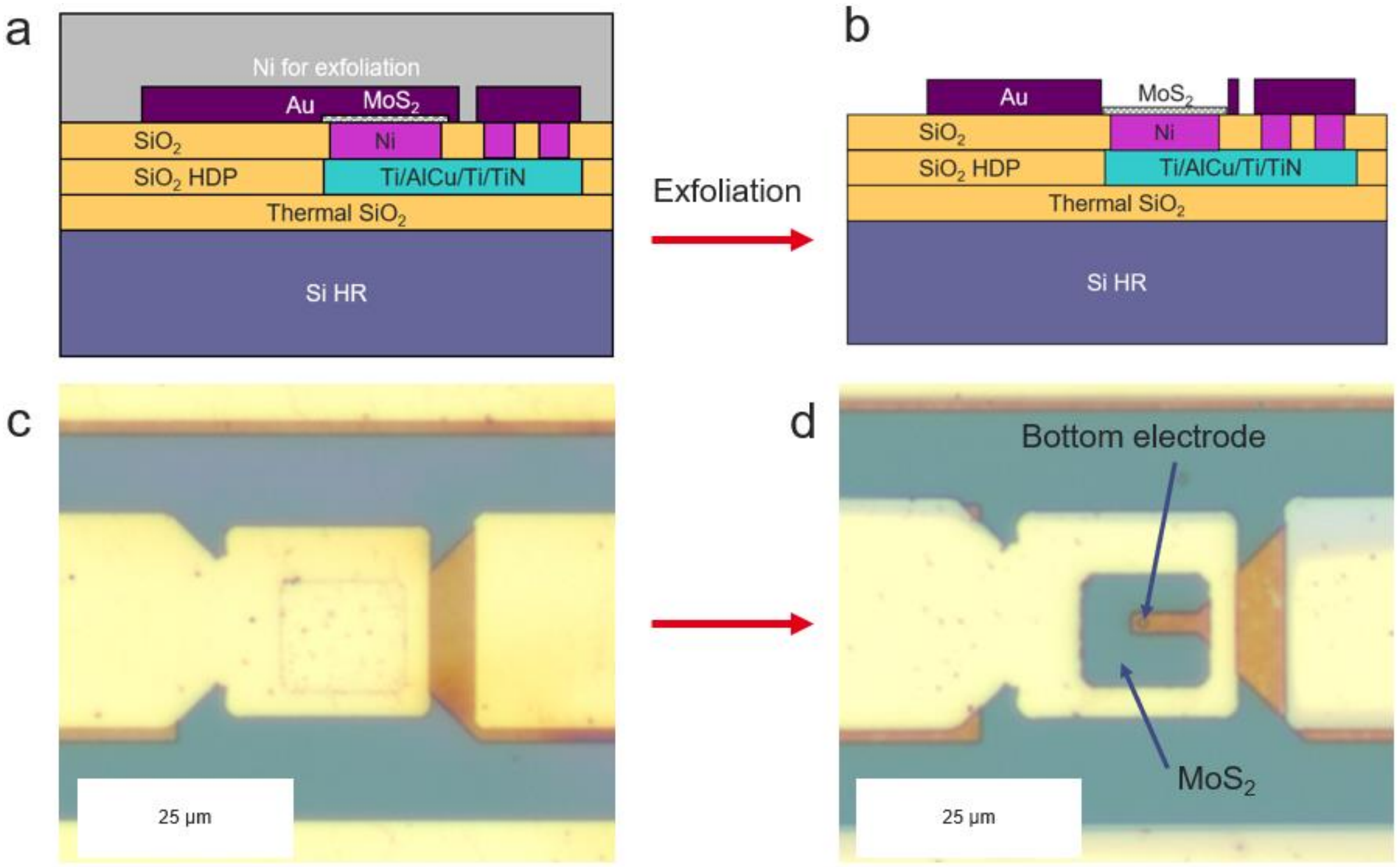

*Figure 3: Illustration of a device before (a) and after (b) the exfoliation of the top electrode. Optical microscopy image in top view of a device before (c) and after (d) the exfoliation of the top electrode.*

## Physico-chemical characterization

After top electrode exfoliation, three different methods are used to identify and characterize the conducting filament. The first one is Kelvin Probe Force Microscopy (KPFM). Different devices are prepared in each electrical state: initial state, ON state and OFF state. First, the bottom electrode is localized by Atomic Force Microscopy (AFM) thanks to a topography difference (the bottom electrode being slightly lower than the $SiO_2$ film). Figure 4 (a) and (b) respectively show the AFM image of a device in ON state and in initial state. Red circles correspond to the bottom electrode. Once the bottom electrode is localized, KPFM measurements can be performed above it. Figure 4 (c) shows the KPFM image of the device in ON state. We can see the presence of a bright spot inside the red circle corresponding to the bottom electrode. Figure 4 (d) shows the KPFM image of the device in initial state with the absence of bright spot above the bottom electrode. Devices in OFF state are also measured exhibiting the same absence of bright spot as in devices in initial state. These results suggest the creation of a conducting filament, with a diameter of approximately 120 nm after the set operation to put the memristor in the ON state. Three devices in each state were tested, giving similar results.

In addition, a Raman spectroscopy map was performed on an ON state device. Figure 4 (e) and (f) respectively show the AFM and KPFM image of the ON state device characterized by Raman spectroscopy. A conducting filament is observable on the right side of the bottom electrode (visualized by the red circle). Figure 4 (g) shows a Raman spectroscopy map of the $A_{1g}$ peak position for this ON state device. We can clearly observe a red shift of the $A_{1g}$ peak at the same position than the bright spot observable in KPFM in Figure 4 (f). This red shift of the $A_{1g}$ peak corresponds to a p-type doping in this area [36].

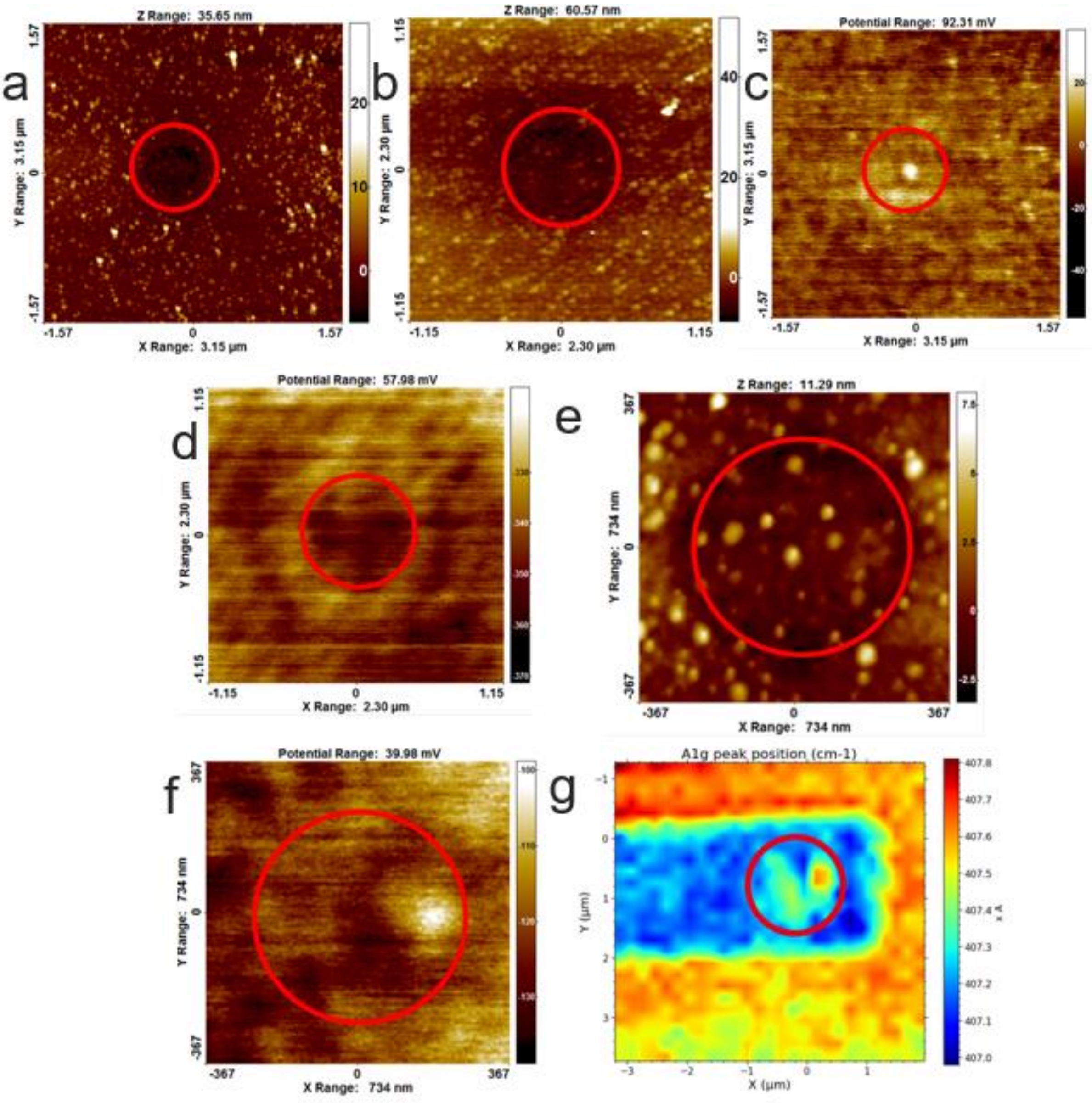

*Figure 4: (a, b) AFM images of an ON state and an initial state respectively (c, d) KPFM images of an ON state and an initial state respectively. (e, f, g) AFM, KPFM and micro-Raman measurements respectively of a device in the ON state.*

To complete this study of the conducting filament, we perform TEM cross sections of devices in initial state, ON state and OFF state. Figure 5 (a) and (b) respectively show a TEM cross section and an illustration of an initial state device. We can observe that gold, $MoS_2$ and nickel are perfectly separated. Figure 5 (c) shows a TEM cross section of an ON state device. Here, we observe a connection between the gold top electrode and the nickel at the right edge of the nickel bottom electrode (red rectangle). Figure 5 (d) is a zoom in the red rectangle of Figure 5 (c). In the blue ellipse, gold has migrated into the $MoS_2$ layer to connect with the nickel electrode. Gold atoms that migrated from the top electrode to the bottom electrode come from the red ellipse where a depletion in gold next to the conducting filament is visible. Figure 5 (e) shows an illustration of the device in ON state where a depletion area is observable with the gold filament crossing the $MoS_2$ layer. Figure 5 (f) is a TEM cross section of a device in OFF state. We can observe the presence of a depletion area in the red ellipse similar to the one observed in Figure 5 (d) but no conducting filament is observable. It suggests that the one formed during the set operation has fully disappeared but that gold has migrated to a different position than the initial one. We thus assume that it migrates at the $MoS_2$/gold interface as illustrated in Figure 5 (g) for the device in OFF state.

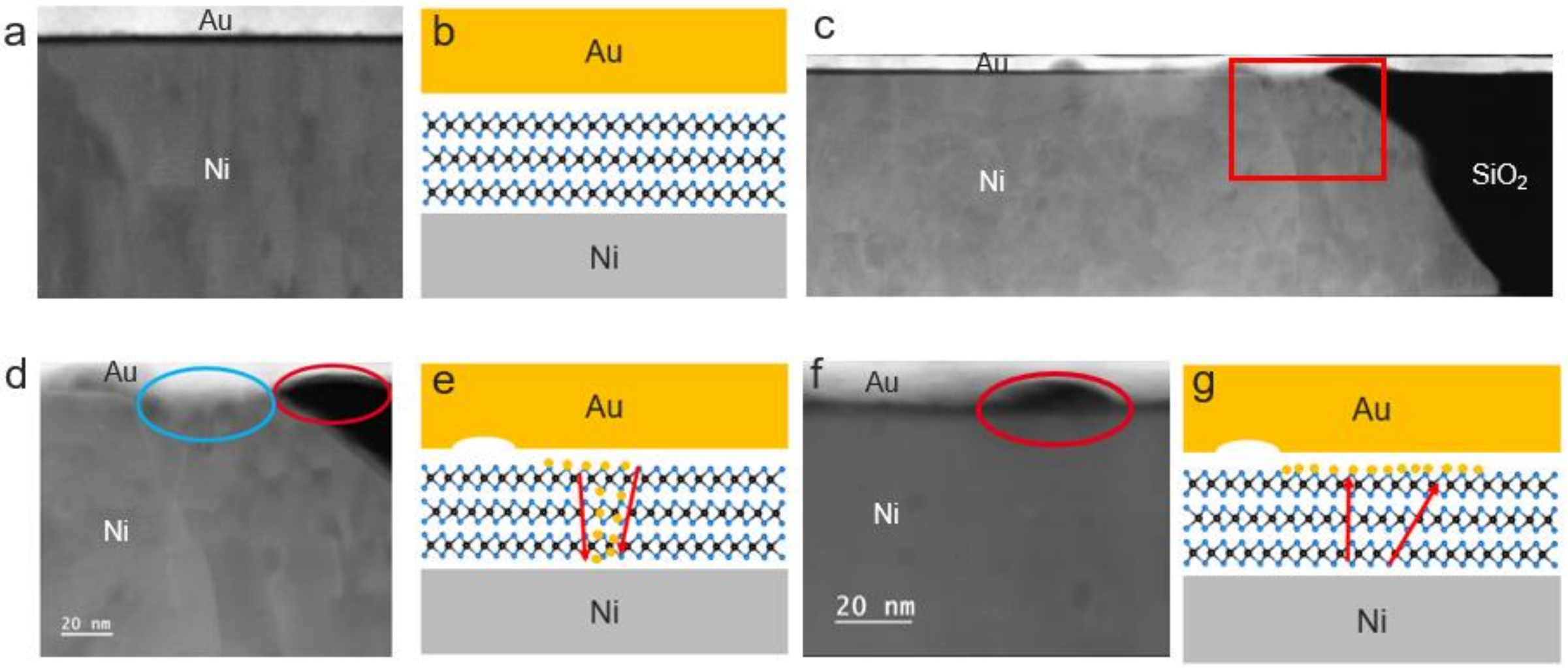

*Figure 5: (a) TEM cross section image of a device in initial state, (b) Illustration of the device in initial state. (c) TEM cross section image of a device in ON state, (d) Zoom in the red square of (a). (e) Illustration of the device in ON state. (f) TEM cross section image of a device in OFF state and (g) Illustration of the device in OFF state.*

## Discussion

The KPFM analysis demonstrates the existence of a conducting path connecting the top and bottom electrodes. Raman spectroscopy also shows a local modification of the $MoS_2$ doping at the same location. These experiments demonstrate that the switching mechanism is due to the formation of a filament. Moreover, the observed doping is p-type which is coherent with a gold migration inside $MoS_2$ [37]. TEM cross-section observations confirmed that this filament is made of gold, the metal of the top electrode.

We can conclude that the switching mechanism in $MoS_2$-based memristors is due to metallic migration from the top electrode into the $MoS_2$ film to form a conducting metallic filament.

Considering this mechanism, the difference in switching parameters between Ni/$MoS_2$/Au and Pt/$MoS_2$/Ni stacks can be explained by the different metal-$MoS_2$ interactions. Taking into account different atomic sites at the $MoS_2$ surface, the adsorption energy of a metallic atom at the $MoS_2$ surface is more than twice larger for a nickel atom (1.7 eV) than for a gold atom (0.66 eV) [38]. Moreover, the diffusion barrier of an adatom between two atomic sites at the surface of the $MoS_2$ is 0.07 eV for gold atoms while it is 0.85 eV for a nickel atom [38]. Thus, the formation of a gold filament in $MoS_2$ requires less energy than a nickel filament. This can explain why the switching voltage is lower for both set and reset operations in Ni/$MoS_2$/Au devices than in Pt/$MoS_2$/Ni devices. Nevertheless, considering that devices tend to be blocked in the conducting state when they stop cycling, we believe that the more favorable creation of a conducting filament also facilitates either the formation of an extra filament or the creation of a filament too thick to be removed by the reset operation. This last point can explain why the endurance is greater for Pt/$MoS_2$/Ni devices in which the formation of a thick filament is a very rare event.

Based on our experimental findings, we propose an explanation for the entire cycling mechanism in $MoS_2$-based memristors. Gold is first pumped from the electrode and adsorbed at the $MoS_2$ surface during the forming operation (Figure 6 (a) and (b)), then it diffuses into the $MoS_2$ to form a filament Figure 6 (c) to complete the forming operation or for the set operation. The reset operation (Figure 6 (d)) is due to the diffusion of metallic atoms from the conducting filament to the $MoS_2$/metal interface. A failure occurs during a set operation when the diffusion mechanism creates a filament too thick (or possibly too many filaments) to be removed by the reset operation (Figure 6 (e)).

The nature of the metallic electrodes being a critical parameter for $MoS_2$ memristors, the study of a wide variety of metallic electrodes is essential to fully optimize such devices.

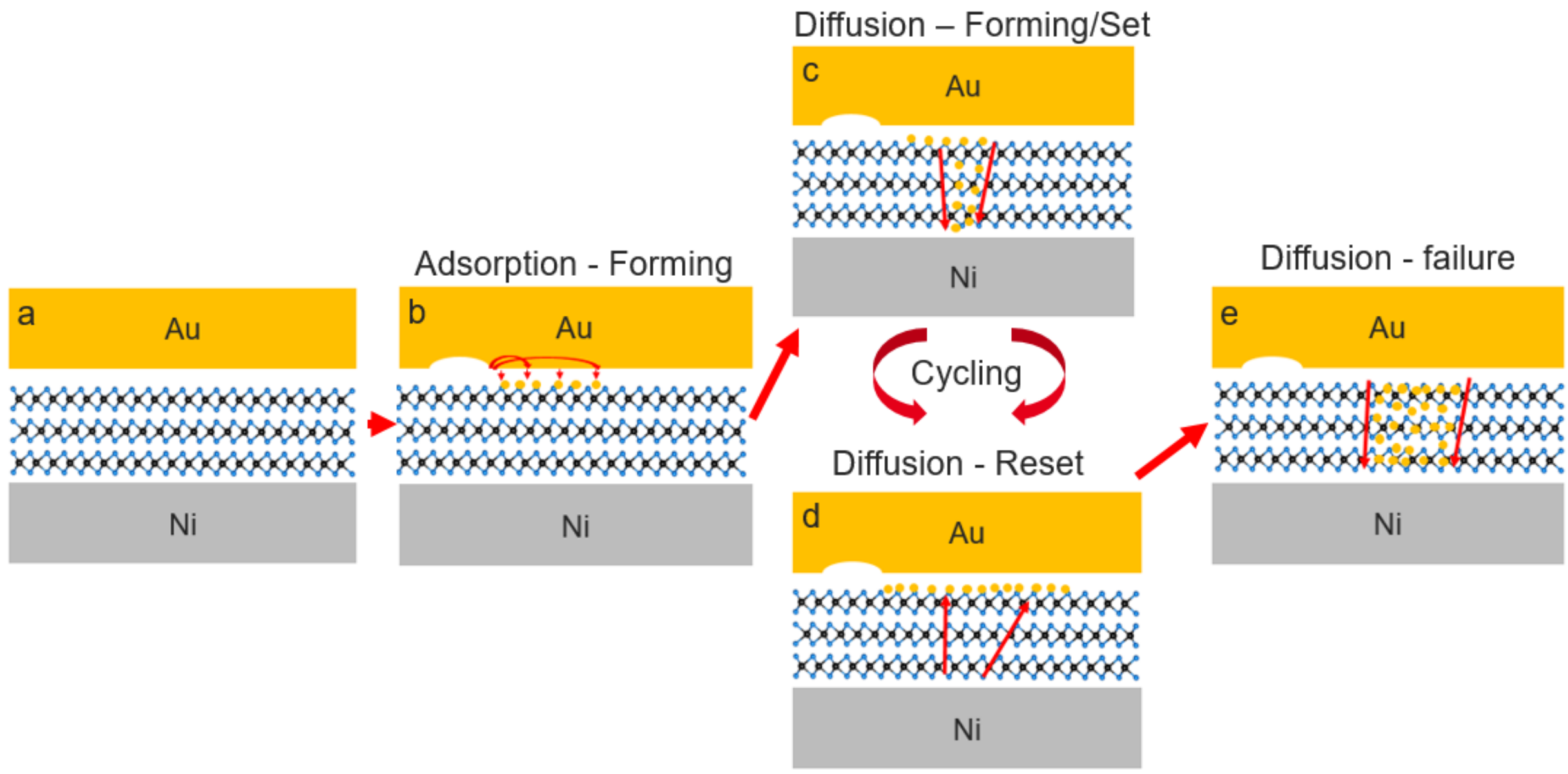

*Figure 6: Illustration of the proposed physical mechanism to explain the commutation of $MoS_2$ based memristors at different steps of cycling: (a) In initial state, (b) metallic adsorption during the forming operation, (c) during the diffusion step for the forming or set operations, (d) during the reset operation and (e) when a failure occurs.*

## Conclusion

In this work, we studied two different memristor stacks by electrical characterization. The Pt/$MoS_2$/Ni stack shows better endurance, reaching more than 100 cycles whereas in Ni/$MoS_2$/Au stacks, only 20 cycles can be reached. In addition, the switching voltage, both for the set and reset operations, is lower for the Ni/$MoS_2$/Au stack. To explain these differences, we unambiguously demonstrate the role played by the metallic electrode in the switching mechanism of 2D material-based memristors. The TEM cross-section view of the device in ON state demonstrates that the conducting filament is formed by the migration of metallic atoms coming from the top electrode.

To complete this observation, an original method allowing to exfoliate the top electrode leaving the $MoS_2$ on the bottom electrode is developed and used to directly characterize the filament with surface characterization tools like KPFM and Raman spectroscopy. The exfoliation of the top electrode to characterize $MoS_2$ memristors is a key progress to characterize the conducting filament and get better insight in the switching mechanism. It allowed us to directly characterize the surface of $MoS_2$ after electrical operations in a working device. Moreover, this top electrode mechanical exfoliation being highly reproducible, the characterization can be possibly performed on a great number of devices which paves the way to statistical studies on the conducting filament that are more difficult to perform with a single characterization technique such as TEM cross section.

Using a full set of characterization techniques, we could identify the conducting filament by observing a doped area corresponding to the gold migration into $MoS_2$. By clarifying the role of metallic electrodes in the switching mechanism, we conclude that the difference in switching behavior between Ni/$MoS_2$/Au and Pt/$MoS_2$/Ni stacks is due to the lower adsorption and diffusion energies of gold atoms into $MoS_2$, making easier the creation of a gold filament than a nickel one. It can explain the lower switching voltage of Ni/$MoS_2$/Au devices and their poorer endurance as due to the formation of extra filaments or filaments too thick to be removed by the reset operation. This work constitutes a real advance in the study of 2D material-based memristors by proposing to a new type of in-operando characterization technique of the switching mechanism.

## Experimental methods

The Amplitude Modulation Kelvin Probe Force Microscopy (AM-KPFM) measurements were performed in a glove box (Argon, ppm $O_2$/$H_2O$) using a Bruker Dimension ICON AFM tool at atmospheric pressure. The topography is recorded in the tapping mode with a free amplitude of 15 nm and an amplitude ratio of 0.8. Then a 15 nm lift height distance is used for the KPFM measurement with a voltage modulation of 500 mV.

Electrical measurements are performed using a HP4156 Semiconductor Analyzer with multiple Source Measurement Units (SMU).

Micro-Raman Cartography is performed with a Spectrometer Raman Nano Soleil from Horiba with a 532 nm wavelength laser.